\documentclass[11pt,twoside]{article}
\usepackage{graphicx}
\usepackage{amsmath}
\usepackage{amssymb}
\usepackage{amscd}
\usepackage{cite}

\usepackage{pstricks}

\begin{document}
\newcommand{\pst}{\hspace*{1.5em}}


\newcommand{\be}{\begin{equation}}
\newcommand{\ee}{\end{equation}}
\newcommand{\bm}{\boldmath}
\newcommand{\ds}{\displaystyle}
\newcommand{\bea}{\begin{eqnarray}}
\newcommand{\eea}{\end{eqnarray}}
\newcommand{\ba}{\begin{array}}
\newcommand{\ea}{\end{array}}
\newcommand{\arcsinh}{\mathop{\rm arcsinh}\nolimits}
\newcommand{\arctanh}{\mathop{\rm arctanh}\nolimits}
\newcommand{\bc}{\begin{center}}
\newcommand{\ec}{\end{center}}

\thispagestyle{plain}

\label{sh}


\begin{center} {\Large \bf
\begin{tabular}{c}
CHAINED QUANTUM ARNOLD TRANSFORMATIONS
\end{tabular}
 } \end{center}

\bigskip

\bigskip

\begin{center} {\bf
Francisco F. L\'opez-Ruiz$^{1*}$, Julio Guerrero$^2$ and Victor Aldaya$^1$
}\end{center}

\medskip

\begin{center}
{\it
$^1$ Instituto de Astrof\'\i sica de Andaluc\'\i a, CSIC, 
\\
  Apartado Postal 3004, 18080 Granada, Spain.

\smallskip

$^2$Departamento de Matem\'atica Aplicada, Universidad de Murcia,
\\
Campus de Espinardo, 30100 Murcia, Spain.
}
\smallskip

$^*$Corresponding author e-mail:~~~flopez~@~iaa.es\\
\end{center}

\begin{abstract}\noindent
We put forward the concatenation of Quantum Arnold Transformations
as a tool to obtain the wave function of a particle subjected to a
harmonic potential which is switched on and off successively. This simulates the
capture and release process of an ion in a trap and provides a mathematical
picture of this physical process.
\end{abstract}

\medskip


\section{Introduction}
\label{intro}
\pst

In a recent paper \cite{QAT} the authors proposed a transformation that maps
states and operators from quantum systems in one dimension described by
generic, time-dependent, quadratic Hamiltonians to the system of the free
particle. This transformation is the quantum version of the Arnold transform
(QAT), which in its original classical version \cite{Arnold} maps solutions of a
certain type of classical equation of motion, a non-homogeneous linear second
order ordinary differential equation (LSODE), to solutions of the classical
equation for the free particle. Through this transformation, the class of
quantum systems with quadratic, time-dependent Hamiltonian can be connected and
its solutions and relevant operators related by using the free particle system
as an intermediate point. This way, the symmetry properties of the free particle
system are transferred to this whole class of quantum systems. 

In fact, the QAT proves to be useful in the construction of wave
functions, the quantum propagator or the evolution operator, taking advantage of
the simple properties of the free particle quantum system. Even more, some
insight in the physics of the free particle can also be obtained by importing
to this system features that are, in principle, naturally attributed to the
quantum harmonic oscillator. This is the case of coherent and squeezed states in
one spatial dimension, as well as Hermite-Gauss and Laguerre-Gauss states in
higher dimensions, as it was shown in \cite{discretebasis}. 

There are some related strategies in the literature employed to solve
the Schr\"odinger equation with a general quadratic, time-dependent Hamiltonian.
We could mention the method of looking for integrals of motion showed, for
instance, in \cite{Manko} and \cite{Suslov-AnPhys,Suslov-Lopez}. Some other
references may be found in \cite{QAT,discretebasis}.

A particular case of time-dependent potential is that of a harmonic potential
which is turned on and off successively, preserving or not the original
frequency $\omega$ of the oscillator. 
This could be useful to simulate the processes of capturing or releasing ions
in traps. Usually, ion traps, like the Penning of the Paul trap, have time
varying, periodic frequencies (for instance, for the Paul trap the classical
equation of motion is a Matieu equation), but we shall restrict ourselves to
constant frequency to focus on the idea to turning on and off the potential
without introducing unnecessary technicalities. 
The purpose of this paper is to take
advantage of the QAT that relates the harmonic oscillator and the free particle
to obtain a method to compute analytically the wave function and the evolution
operator for such a system.

The complete time evolution operator along a lapse of time in which a
particle is released and captured several times in a harmonic potential will be
a product of evolution operators for the free particle and the harmonic
oscillator. This is a consequence of the sudden or diabatic approximation if
the processes of switching on (and off) are fast enough. The harmonic-oscillator
evolution operation on certain states
(namely eigenstates of the harmonic oscillator Hamiltonian) may be simpler than
the free one. It will be shown that it is possible to split each free time
evolution operator in two operations: a harmonic evolution and a quantum Arnold
transformation, in such a way that the complete evolution can be written in
terms of harmonic evolutions and QATs only. This might not be always the easiest
way to compute the resulting wave function. We believe nevertheless that it is
interesting to have the possibility that we present here at our disposal. This
technique has also the advantage of being easily generalizable to generic
quadratic potentials with abrupt jumps in its time-dependence. However, we will
keep the simplicity of the harmonic potential to keep the idea as neat as
possible.

The paper is organized as follows. In Sec.~\ref{theqat} we give an overview of
the Quantum Arnold Transformation for general quadratic Hamiltonians.
In Sec.~\ref{harmonic-free} we provide the particularization for the case of
the harmonic oscillator and fix some notation. In Sec.~\ref{switchonoff} we
obtain the alternative expression for the time evolution operator in terms of
QATs. A summary and some comments are given in Sec.~\ref{summary}.

\section{The Quantum Arnold Transformation}
\label{theqat}
\pst

The quantum Arnold transformation (or the inverse) relates the Hilbert space
$\mathcal H_t$ of solutions of the free Schr\"odinger equation 
\begin{equation}
i\hbar\frac{\partial \varphi}{\partial t}=-\frac{\hbar^{2}}{2m}
\frac{\partial^{2} \varphi}{\partial x^{2}}\, ,
\label{eq:Schrodinger libre}
\end{equation}
associated with the corresponding classical equation
\begin{equation}
\ddot{x}=0\,,
\label{eq:ecuacion libre}
\end{equation}
to that space $\mathcal H'_{t'}$ of solutions of the Schr\"odinger equation
\begin{equation}
i\hbar\frac{\partial \phi'}{\partial t'}=-\frac{\hbar^{2}}{2m}e^{-f}
\frac{\partial^{2} \phi'}{\partial x'^{2}}+\bigl(\frac{1}{2}m\omega^{2}x'^{2}
\bigr)e^{f}\phi' \,,
\label{eq:Schrodinger general}
\end{equation}
where the quantum theory of a generic LSODE\footnote{We consider here the case
of homogeneous classical equation and the corresponding quantum theory. For the
presence of an external force, see \cite{QAT}.}
\begin{equation}
\ddot{x}'+\dot{f}\dot{x}'+\omega^{2}x'= 0
\label{eq:ecuacion de movimiento general}
\end{equation}
is realized. The quantities $f$ and $\omega$ are, in general, dependent on
time $t'$.
The classical equation (\ref{eq:ecuacion de movimiento general}) can be derived
from a Hamiltonian function
\begin{equation}
H=\frac{p'^{2}}{2m}e^{-f}+\bigl(\frac{1}{2}m\omega^{2}x'^{2} \bigr)e^{f}\,,
\label{eq:hamiltoniano general}
\end{equation}
and, according to the standard canonical prescriptions, it leads to the
Schr\"odinger equation (\ref{eq:Schrodinger general}).
For $f$ linear in time and constant $\omega$
this equation is commonly known as Caldirola-Kanai equation for the damped
harmonic oscillator \cite{Caldirola,Kanai}.

Being both spaces of solutions of (\ref{eq:Schrodinger libre}) and
(\ref{eq:Schrodinger general}), $\mathcal H_t$ and
$\mathcal H'_{t'}$ respectively, related, we will also have the basic, constant
of motion quantum
operators associated with the classical functions position and momentum
connected and realized as well-defined, constant of motion operators on both
$\mathcal H_t$ and $\mathcal H'_{t'}$.
However, it is important to note that the Hamiltonian of one of the systems is
\textit{not} transformed into the Hamiltonian of the other: the
generators of the symmetry of the free particle, the Schr\"odinger group, are
mapped into a set of operators closing the same Lie algebra in the system
corresponding to the generic LSODE. The Hamiltonian operator of the LSODE-system
is not the same abstract element in this algebra or, even, it does
not necessarily belongs to this symmetry algebra. 

This is a generalization of the classical Arnold transformation $A$, and is
obtained by completing $A$ with a change of the wave function. While the
classical Arnold transformation $A$ is given by:
\begin{equation}
\begin{split}
  A:& \; \; \mathbb R \times T' \longrightarrow \mathbb R \times T \\
    & \quad (x',t')  \longmapsto \; (x,t) =
        A\bigl((x',t')\bigr) = (\tfrac{x'}{u_2},\tfrac{u_1}{u_2})\,,
\end{split}
\end{equation}
where $T'$ and $T$ are open intervals of the real line containing $t'=0$
and $t=0$, respectively, and $u_1(t')$ and $u_2(t')$ are independent solutions
of the classical equation of motion in $(x',t')$ (\ref{eq:ecuacion de movimiento
general}), the QAT is written:
\begin{equation}
\begin{split}
  \hat A:& \quad \mathcal H'_{t'} \;\; \longrightarrow \quad \mathcal H_t \\
    & \phi'(x',t')  \longmapsto \; \varphi(x,t) = 
        \hat A \bigl( \phi'(x',t') \bigr) = 
       A^* \bigl( \sqrt{u_{2}(t')}\,e^{-\frac{i}{2}\frac{m}{\hbar}
                 \frac{1}{W(t')}\frac{\dot{u}_{2}(t')}{u_{2}(t')}
                                                 x'^{2}} \phi'(x',t') \bigr)
\,,
\end{split}
\end{equation}
where $A^*$ denotes the pullback operation corresponding to $A$, and
$W(t') \equiv \dot{u}_{1} u_{2} - u_{1} \dot{u}_{2}=e^{-f}$. It is
straightforward to check that by this transformation the Schr\"odinger equation
of the free particle is transformed into (\ref{eq:Schrodinger general}) up
to a multiplicative factor which depends on the particular choice of the
classical solutions $u_1$ and $u_2$ (partial derivatives must be changed by the
classical part of the transformation while wave functions are shifted by the
quantum part).

We impose on $u_1$ and $u_2$ the condition that they preserve the
identity of $t$ and $x$, i.e., that $(x,t)$ coincide with $(x',t')$ at an
initial point $t'_0$, arbitrarily taken to be $t'_0=0$:
\begin{equation}
 u_1(0)=0, \; u_2(0)=1, \qquad \dot{u}_1(0)=1, \; \dot{u}_2(0)=0 \,. 
\label{eq:condiciones} 
\end{equation}
This fixes a unique form of $A$ for a given ``target'' LSODE-type physical
system. However, the quantum Arnold transformation is still valid if solutions
$u_1$ and $u_2$ do not satisfy \eqref{eq:condiciones} (see \cite{QAT} for
details).

It will be useful to have a pictorial representation of the situation. There
will be a common Hilbert space $\mathcal H$ of wave functions, which plays the
role of initial values for both the solutions $\phi'(x',t') \in \mathcal
H'_{t'}$ and $\varphi(x,t) \in \mathcal H_t$; evolution operators, $\hat
U(t,0)\equiv \hat U(t)$ and $\hat U'(t',0)\equiv\hat U'(t')$, acting on the
initial wave functions; and the quantum Arnold transformation $\hat A$
that relates Schr\"odinger equations and basic operators:  
\begin{equation}
\begin{CD} 
\mathcal H_t @<\hat A<< \mathcal H'_{t'}\\ 
@A{\hat U(t)}AA @AA\hat U'(t')A\\ 
\mathcal H_0 \equiv \mathcal H @>>\hat{1}> \mathcal H \equiv \mathcal H'_0 
\end{CD}
\label{diagram}
\end{equation}
where $\hat 1$ is the identity operator. Had the solutions $u_1$ and $u_2$
not been chosen satisfying (\ref{eq:condiciones}), the price would have been
that the relation in the lower part of the diagram \eqref{diagram} above would
no longer be the identity and basic position and momentum operators would then
be mixed (see \cite{QAT}).

\bigskip

Let us stress that QAT can be useful to quickly perform some
calculations, avoiding tedious, direct evaluations which can become extremely
involved in the system under study. For example, it can be used to compute
the quantum propagator for any LSODE-type quantum system, following the idea of
Takagi in \cite{Takagi} for the simple case of the harmonic oscillator, or even
the evolution operator $\hat U' (t')$, which becomes very difficult to evaluate
exactly when the Hamiltonian is time-dependent and does not commute with itself
at different times. 

Actually, the evolution operator of a LSODE system can be related with the free
evolution operator. Having in mind the diagram \eqref{diagram}, we write:
\begin{equation}
  \hat A \bigl( \hat U' (t') \phi' (x')\bigr) = \hat U (t) \varphi (x)
\,.
\end{equation}
Here $\phi'$ and $\varphi$ are \textit{the same} function of only one argument
($x$ or $x'$) and we will denote $\varphi = \phi' = \psi$. Then,
\begin{multline}
  \hat U' (t') \psi(x') = \hat A^{-1} \bigl(\hat U (t) \psi (x)
\bigr) 
= 
   \frac{1}{\sqrt{u_{2}}} e^{\frac{i}{2}\frac{m}{\hbar}
                 \frac{1}{W}\frac{\dot{u}_{2}}{u_{2}} x'^{2}} 
    A^{*-1} \bigl(\hat U (t) \psi (x) \bigr) 
= \\ =
  \frac{1}{\sqrt{u_{2}}} e^{\frac{i}{2}\frac{m}{\hbar}
                 \frac{1}{W}\frac{\dot{u}_{2}}{u_{2}} x'^{2}} 
    A^{*-1} \bigl(\hat U (t) \bigr) A^{*-1} \bigl(\psi (x) \bigr) 
\,.
\end{multline}
To factorize the function $\psi$ and single out the general action of 
$\hat U' (t')$, we compute
\begin{equation}
  A^{*-1} \bigl(\psi (x) \bigr) = \psi (\tfrac{x'}{u_2}) = 
      e^{\log(1/u_2) x' \frac{\partial}{\partial x'}}\psi (x') \,,
\end{equation}
where $ e^{\log(1/u_2) x' \frac{\partial}{\partial x'}}$ is a dilation operator
which is \textit{not} unitary. To unitarize this operator, the generator must be
shifted (symmetrized) from $x' \frac{\partial}{\partial x'}$ to $x'
\frac{\partial}{\partial
x'} + \frac{1}{2}$, so that the true unitary operator is then 
\begin{equation}
 \hat U_D (\tfrac{1}{u_2}) = e^{\log(1/u_2) (x' \frac{\partial}{\partial
x'} + \frac{1}{2})} = \frac{1}{\sqrt{u_2}} e^{\log(1/u_2) x'
\frac{\partial}{\partial x'}} \,.
\end{equation}
But the factor $\frac{1}{\sqrt{u_2}}$ is already present in the previous
expression of $\hat U' (t')$. Therefore, it now reads
\begin{multline}
  \hat U' (t') =  
       e^{\frac{i}{2} \frac{m}{\hbar}\frac{1}{W}
           \frac{\dot u_2}{u_2} x'^2} 
          A^{*-1} \bigl( \hat U (t) \bigr) \hat U_D (\tfrac{1}{u_2}) 
= \\ =
      \frac{1}{\sqrt{u_2}} e^{\frac{i}{2} \frac{m}{\hbar}\frac{1}{W}
           \frac{\dot u_2}{u_2} x'^2} 
      e^{\frac{i \hbar}{2m} u_1 u_2 \frac{\partial^2}{\partial x'^2}}
      e^{\log(1/u_2) x' \frac{\partial}{\partial x'}}\, .
\qquad \qquad \qquad
\label{eq:operador evolucion general}
\end{multline}
%

It is remarkable that we have been able to write an \textit{exact} expression
for the evolution operator as a product of operators. No perturbative
approximation method, which could become cumbersome in some cases, is needed for
\textit{any} LSODE-related quantum system to obtain the evolution operator.

\section{A particular case: the harmonic oscillator and the free particle}
\label{harmonic-free}
\pst

Now let us give the specific expressions for the harmonic oscillator,
that is to say, $\dot{f}=0$ and $\omega(t')=\omega$, constant. The two
independent classical solutions can be chosen as 
\begin{equation*}
  \begin{split}
u_1(t')&=\frac{1}{\omega}\sin(\omega t')    
\\
u_2(t')&=\cos(\omega t')
  \end{split}\,,
\end{equation*}
with $W(t')=1$. It can be checked that the change of variables in the classical
Arnold transformation results in:
\begin{equation*}
  \begin{split}
 t'&=\frac{1}{\omega}\tan^{-1}(\omega t)
\\
 x'&=\cos(\tan^{-1}(\omega t))x=\frac{x}{\sqrt{1+\omega^2 t^2}}
\label{ArnoldClasico}
  \end{split}\,.
\end{equation*}
The solutions $u_1$ and $u_2$ satisfy the conditions (\ref{eq:condiciones}).
Therefore, the diagram (\ref{diagram}) can be simplified to:

\begin{equation}
\scalebox{1} 
{
\begin{pspicture}(0,-1.8492187)(7.4490623,1.8492187)
\psline[linewidth=0.02cm,arrowsize=0.05291667cm 2.0,
arrowlength=1.4,arrowinset=0.4]{<-}(1.77,1.1157813)(4.37,1.1157813)
\usefont{T1}{ptm}{m}{n}
\rput(0.9245312,1.1057812){${\cal H}$}
\usefont{T1}{ptm}{m}{n}
\rput(5.5045314,1.0857812){${\cal H}_{\rm HO}$}
\psline[linewidth=0.02cm,arrowsize=0.05291667cm 2.0,
arrowlength=1.4,arrowinset=0.4]{<-}(1.11,0.55578125)(2.59,-1.2842188)
\psline[linewidth=0.02cm,arrowsize=0.05291667cm 2.0,
arrowlength=1.4,arrowinset=0.4]{<-}(4.93,0.55578125)(3.33,-1.2842188)
\usefont{T1}{ptm}{m}{n}
\rput(3.0745313,-1.6142187){${\cal H}_{\rm 0}$}
\usefont{T1}{ptm}{m}{n}
\rput(0.8645313,-0.37421876){$\hat{U}$}
\usefont{T1}{ptm}{m}{n}
\rput(5.204531,-0.35421875){$\hat{U}^{\rm HO}$}
\usefont{T1}{ptm}{m}{n}
\rput(3.2445312,1.6457813){$\hat{A}$}
\end{pspicture} 
}
\label{triangle}
\end{equation}
%
%
%
%
%
where $\hat{U}$ and $\hat{U}_{\rm HO}$ stand for the evolution operators of the
free particle and the harmonic oscillator, respectively, while $\mathcal  H_0$
is the Hilbert space, either for the free particle or the harmonic oscillator,
of solutions of their respective Schr\"odinger equations at $t=0$.

It is possible to apply the QAT, for instance, to the time-dependent eigenstates
of the harmonic oscillator Hamiltonian $\hat H_{\rm HO}'$, with energy
$E_n=\hbar \omega(n+\frac{1}{2})$ and belonging to ${\cal H}_{\rm HO}$,
\begin{equation}
\psi'_n(x',t') ={\cal N}_n
\hbox{\Large \it e}^{-i\omega(n+\frac{1}{2}) t'}
\hbox{\Large \it e}^{-\frac{m\omega}{2\hbar}x'^2}
H_n(\sqrt{\frac{m\omega}{\hbar}} x' )
\,,\label{eigenHO}
\end{equation}
where ${\cal N}_n=\left(\frac{m \omega }{\hbar \pi}\right)^{\frac{1}{4}
}\frac{1}{\sqrt{2^n n!} }$. We obtain the following set of states, solutions of
the Schr\"odinger equation for the free particle:
\begin{equation}
\psi_n(x,t) ={\cal N}_n\frac{1}{\sqrt{|\delta|}}
\hbox{\Large \it e}^{-\dfrac{x^2\delta^*}{4L^ 2 |\delta|^2}}
\left(\frac{\delta^*}{|\delta|}\right)^{n+\frac{1}{2}}H_n(\frac{x}{\sqrt{2}
L|\delta| } )\,,
\label{discretebasis}
\end{equation}
where, in order to obtain a more compact notation, we have introduced the
quantities $L=\sqrt{\frac{\hbar}{2m\omega}}$, with dimensions of length,
and $\tau=\frac{2mL^ 2}{\hbar}=\omega^{-1}$, with dimensions of time. 
We also denote by $\delta$ the complex, time dependent, dimensionless
expression $\delta=1+i\omega t=1+ i\frac{\hbar t}{2mL^2}=1+i \frac{t}{\tau}$.
We have also used the fact that $\hbox{\Large \it e}^{-i\omega t'}=
\hbox{\Large \it e}^{-i {\rm tan}^{-1}(\omega t)}=\frac{\delta^*}{|\delta|}$.
Note that with these definitions the normalization factor ${\cal N}_n$ can be 
written as ${\cal N}_n=\frac{(2\pi)^{-\frac{1}{4}}}{\sqrt{2^n n! L}}$.



This set of states constitutes a basis for the space of solutions of the
free Schr\"odinger equation, since it is mapped from a basis for the harmonic
oscillator through $\hat A$, which is unitary. 
The first state of this basis, the one mapped from the harmonic oscillator
vacuum state, is given by:
\begin{equation}
\psi_0(x,t) 
   =
\frac{(2\pi)^{-\frac{1}{4}}}{\sqrt{L|\delta|}}
\left(\frac{\delta^*}{|\delta|}\right)^\frac{1}{2}
                    \hbox{\Large \it e}^{-\dfrac{x^2\delta^*}{4L^ 2|\delta|^2}}
		     =\frac{(2\pi)^{-\frac{1}{4}}}{\sqrt{L\delta}}
                     \hbox{\Large \it e}^{-\dfrac{x^2}{4L^ 2 \delta}}\,,
		     \label{paquete-gaussiano}
\end{equation}
which is nothing other than a Gaussian wave packet with center at the origin and
width $L$. The parameter $\tau$ is the dispersion time of the Gaussian wave
packet (see, for instance, \cite{galindo}). The $n$-th state represents a wave
packet with $n+1$ humps (see \cite{discretebasis}). 


The family of wave functions (\ref{discretebasis}) has been known in the
literature as Hermite-Gauss wave packets \cite{Andrews}, and they have been
widely used, in their two dimensional version, in paraxial wave optics 
\cite{Fotonics}.
Note that, making use of the classical solutions only, and through the QAT, we
have been able to import the time evolution from the stationary states of the
harmonic oscillator, $\psi'_n(x',t')$, to the non-stationary ones,
$\psi_n(x,t)$, without solving the time-dependent Schr\"odinger equation. 
This will be the motivation for the next section.


\section{Switching on and off the harmonic potential}
\label{switchonoff}
\pst
Time evolution of an eigenstate of a given Hamiltonian, and in
particular the harmonic oscillator, is encoded in just a time-dependent phase
factor in the wave function. We have seen that if a physical system $S$,
defined by another Hamiltonian, can be linked to the harmonic oscillator through
an Arnold transformation, then the quantum time evolution of $S$ can be
found performing a change of variables in the wave function of the harmonic
oscillator and multiplying by a specific phase and factor, that is, carrying out
the QAT, provided the necessary classical solutions are known. This way, it is
possible to avoid solving the Schr\"odinger equation or computing the evolution
operator whenever the initial state of the system $S$ is an eigenstate of the
harmonic oscillator Hamiltonian. The same observation can be made if we have
knowledge of the time evolution of a certain state of the harmonic oscillator
and we use it as a initial state of $S$. To illustrate the following reasoning,
we are going to ascribe to the free particle system the role of $S$ and use the
results of the previous Section. 

The physical situation we intend to describe is that of a particle in a harmonic
potential which is switched off at a given time $T_0$. After that, the particle
will evolve freely and then, at time $T_1$, is captured again by another (in
particular, the same) harmonic potential. Time evolution will be harmonic again
until it is released once more at $T_2$. Finally, it will be detected at time
$T_3$. 

This kind of process is of practical relevance, for instance in the preparation
of the 1-dimensional Hermite-Gauss free states (\ref{discretebasis}). This idea
was proposed in \cite{QuantumSling}, and was named ``Quamtum Sling''. The vacuum
state of the harmonic oscillator, when switched off, will provide the ``vacuum''
Gaussian wave packet with width $L=\sqrt{\frac{\hbar}{2m\omega}}$, where $m$ is
the mass of the particle and $\omega$ the frequency of the oscillator. Note that
the dispersion time $\tau$ coincides with the inverse of the frequency of the
oscillator. If the harmonic oscillator is in the $n$-th excited state, then a
state with $(n+1)$ ``humps'' is obtained.

If we ``capture'' one of these traveling states at time $T_1$ switching on a
harmonic oscillator potential with an appropriate frequency $\omega_1$, it
would ``freeze'' in a harmonic oscillator state (without
dispersion), until the potential is switched off again at a time $T_2$
and the wave packet is released, traveling again as a free wave packet that
disperses in time. This way, information might be stored temporally in this
``oscillator traps'', which can also be used to further manipulating them or
even measuring the resulting state by means of adequate lasers. The frequency
$\omega_1$ required to capture the dispersed wave packet might be fine tuned in
such a way that the wave packet, at the time $T_1$, matches an appropriate
eigenstate (with the same $n$) of the harmonic oscillator with frequency
$\omega_1$ up to a  phase (see below). 

Which is the resulting wave function $\psi(T_3)$ after the whole process, given
a state prepared for the initial harmonic oscillator $\psi'(T_0)$? Obviously,
the formal solution will be a product of three evolution operators, describing
free evolution from $T_0$ to $T_1$, harmonic with new frequency $\omega_1$ from
$T_1$ to $T_2$ and free from $T_2$ to $T_3$: 
\begin{equation}
\psi(T_3) 
=\hat{U}(T_3,T_2)\hat{U}'_{\omega_1}(T_2,T_1) \hat{U}(T_1,T_0)\psi'(T_0) \,.
\label{eq:evolucionnaif} 
\end{equation}

Before proceeding, let us discuss a little bit of notation. In the previous
sections, we have used lowercase, unprimed letters for quantities referring
to the free particle and lowercase, primed ones for those of the harmonic
oscillator. Now we are indicating \textit{physical} or \textit{true} time with
capital $T$'s. As the Arnold transformation includes a diffeomorphism in time,
if we are going to use it as a tool to perform calculations, we have to be very
careful of not confusing the \textit{physical} time with that used in the Arnold
transformations. Recovering the notation of lowercase letters, fix
$T_0=t_0=t'_0$. Then 
%
%
$T_1 = t_1 = \frac{u_1(t'_1)}{u_2(t'_1)}$, with $u_1$ and $u_2$ satisfying
(\ref{eq:condiciones}) at $t'_0$, and $t'_2=T_2$, for each evolution to be the
initial condition of the next one. Let us also call $t=T_3$. 

Regarding the transformations we are going to use, denote by
$\hat{A}_{\omega,t_0}(t)$ the QAT from a harmonic oscillator of frequency
$\omega$ performed at time $t$, and by $\hat{U}'_\omega$ the unitary time
evolution operator for that harmonic oscillator. With  this notation and the
previous choice of classical solutions $u_1(t')$ and $u_2(t')$, for $t = t_0$
$\hat{A}_{\omega,t_0}(t_0)$ is the identity.

Now, the product (\ref{eq:evolucionnaif}) can be decomposed by a sequence of
QAT's and harmonic oscillator evolution operators splitting each free
evolution operator as the figure (\ref{triangle}) in Sec.~\ref{harmonic-free}
suggests: 
\begin{equation}
\begin{split}
\psi(t) 
&= \hat{A}_{\omega_1,t_2}(t) \hat{U}'_{\omega_1}(t',t_2)
\hat{U}'_{\omega_1}(t_2,t_1)
\hat{A}_{\omega,t_0}(t_1) \hat{U}'_{\omega}(t'_1,t_0)\psi'(t_0) =
\\
&=\hat{A}_{\omega_1,t_2}(t) \hat{U}'_{\omega_1}(t',t_1)
\hat{A}_{\omega,t_0}(t_1) \hat{U}'_{\omega}(t'_1,t_0)\psi'(t_0)\,,
\end{split}
\end{equation}
%
%
%
where $t=\frac{\tilde{u}_1(t')}{\tilde{u}_2(t')}$, with $\tilde{u}_1$ and
$\tilde{u}_2$ satisfying (\ref{eq:condiciones}) at $t_2$. This way, in the last
expression only evolution operators for harmonic oscillators appear. 

The following diagram helps to visualize the setup: 
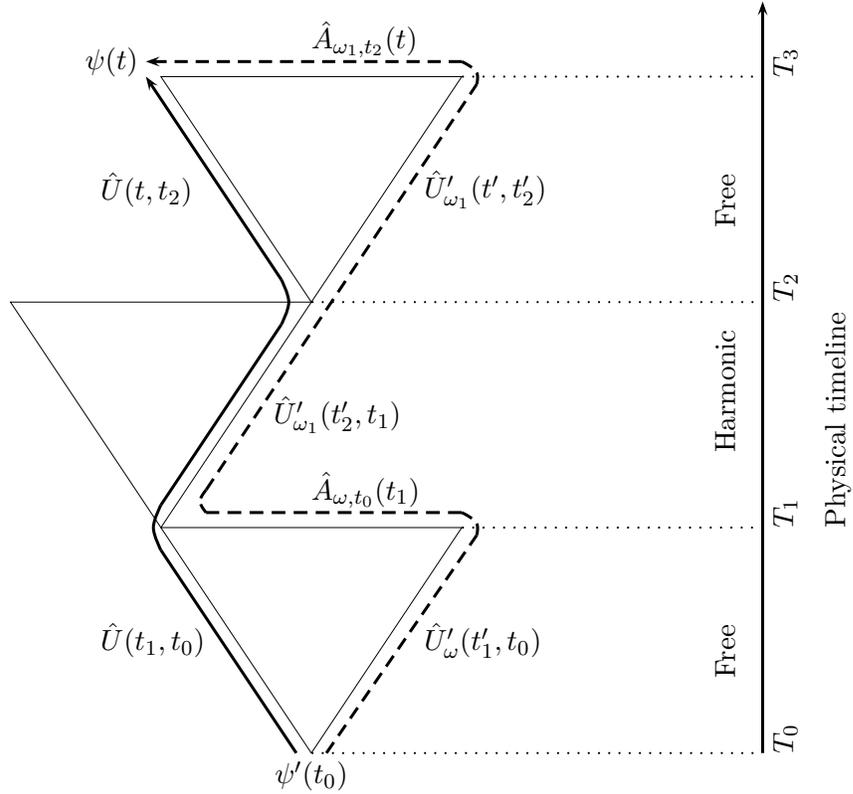
\begin{center}
\scalebox{1} 
{
\begin{pspicture}(0,0)(6,11)
\psline[linewidth=0.01](2,1)(4,4)(0,4)(2,1)
\psline[linewidth=0.01](0,4)(2,7)(-2,7)(0,4)
\psline[linewidth=0.01](2,7)(4,10)(0,10)(2,7)
%
\psline[linewidth=0.04](1.8,1)(0,3.7)
\pscurve[linewidth=0.04](0,3.7)(-0.1,4)(0,4.3)
\psline[linewidth=0.04](0,4.3)(1.6,6.7)
\pscurve[linewidth=0.04](1.6,6.7)(1.7,7)(1.6,7.3)
\psline[linewidth=0.04]{->}(1.6,7.3)(-0.2,10)
%
\psline[linewidth=0.04,linestyle=dashed](2.2,1)(4.2,3.9)
\pscurve[linewidth=0.04,linestyle=dashed](4.2,3.9)(4.2,4.05)(4,4.2)
\psline[linewidth=0.04,linestyle=dashed](4,4.2)(0.6,4.2)
\pscurve[linewidth=0.04,linestyle=dashed](0.6,4.2)(0.52,4.35)(0.6,4.5)
\psline[linewidth=0.04,linestyle=dashed](0.6,4.5)(4.2,9.9)
\pscurve[linewidth=0.04,linestyle=dashed](4.2,9.9)(4.2,10.05)(4,10.2)
\psline[linewidth=0.04,linestyle=dashed]{->}(4,10.2)(-0.2,10.2)
\psline[linewidth=0.04]{->}(8,1)(8,11)
\psline[linewidth=0.03,linestyle=dotted](2,1)(8,1)
\psline[linewidth=0.03,linestyle=dotted](4,4)(8,4)
\psline[linewidth=0.03,linestyle=dotted](2,7)(8,7)
\psline[linewidth=0.03,linestyle=dotted](4,10)(8,10)
\rput[l]{90}(8.3,1){$T_0$}
\rput[l]{90}(8.3,4){$T_1$}
\rput[l]{90}(8.3,7){$T_2$}
\rput[l]{90}(8.3,10){$T_3$}
\rput[l]{90}(9,4){Physical timeline}
\rput[l]{90}(7.5,2){Free}
\rput[l]{90}(7.5,5){Harmonic}
\rput[l]{90}(7.5,8){Free}
\rput[l](-0.8,2.5){$\hat U (t_1,t_0)$}
\rput[l](-0.8,8.5){$\hat U (t,t_2)$}
\rput[l](3.5,2.5){$\hat U'_{\omega} (t'_1,t_0)$}
\rput[l](1.5,5.5){$\hat U'_{\omega_1} (t'_2,t_1)$}
\rput[l](3.5,8.5){$\hat U'_{\omega_1} (t',t'_2)$}
\rput[l](2,4.5){$\hat A_{\omega,t_0}(t_1)$}
\rput[l](2,10.5){$\hat A_{\omega_1,t_2}(t)$}
\rput[b](2,0.5){$\psi'(t_0)$}
\rput[rb](-0.3,10){$\psi(t)$}
\end{pspicture} 
}
\end{center}
Here, the solid line indicates the physical process followed and the dashed 
line the ``computational'' timeline we propose here. The line on the right side
denotes the physical timeline. 

An obvious generalization of the proposed method consists in increasing the
number of captures and releases. On the other hand, depending on the particular
initial state, it may be a better choice a path of calculation along purely free
evolutions and inverse Arnold transformations. 

Note that the evolution will be particularly simple if the frequency $\omega_1$
is such that the width of the wave packet fits a natural width of this second
harmonic potential\footnote{Intuitively, one might say that, in this
situation, the particle would be captured in an eigenstate of the oscillator
Hamiltonian with frequency $\omega_1$. However, this is not the case: it can
be checked that an extra, position-dependent phase appears. This fact and
the relationship with the ``quasistationary'' or ``pseudostationary'' states
\cite{Manko} will be analyzed elsewhere.}. When the frequency of the harmonic
oscillator is not
modified, and the same frequency  $\omega$ is used to capture the state in the
harmonic oscillator trap, the resulting state will be a squeezed state (up
to a phase) with
squeezing parameter $r$ given by $r=-\log(|\delta_1|)$ (where $\delta_1 =
1+i\omega t_1$), which is negative.
This can be seen as a feasible way of producing squeezing in trapped states, 
simply switching off-switching on the trap for a lapse of time $t$, resulting
in a squeezing parameter $r=-\frac{1}{2}\log(1+\omega^2 t^2)$. In fact, a
similar way of producing squeezed states in Bose-Einstein Condensates was
reported in \cite{bec1}.

\section{Summary and comments}
\label{summary}
\pst
We have reviewed the formulation of the quantum Arnold transformation,
particularized it for the relationship between the free particle and harmonic
oscillator and suggested to use a series of QATs and pure harmonic oscillator
evolutions to perform the calculations to obtain the resulting wave function
of a particle that is subjected to the action of turning on an off a harmonic
potential. 

It must be emphasized that this method can be easily generalized to different
quadratic, even time-dependent, potentials and external forces, both changing
their time dependence abruptly. Even in this case, only classical solutions of
each ``phase'' of the potential and external force are required to construct
the resulting wave function and evolution operator. We believe that this tool
may be useful for analytical computations. 

\section*{Acknowledgments}
\pst
Work partially supported by the Spanish MCYT, Junta de Andaluc\'\i a and
Fundaci\'on S\'eneca under projects FIS2008-06078-C03-01,
P06-FQM-01951 and 08816/PI/08.

\end{document}